\documentclass[aps,prb,twocolumn,preprintnumbers,amssymb,showpacs,superscriptaddress,floatfix]{revtex4-1}

\usepackage{amstext,amssymb}
\usepackage{amsmath}
\usepackage{amsfonts}
\usepackage{graphicx}
\usepackage{dcolumn}
\usepackage{bm}
\usepackage{CJK}
\usepackage[dvipsnames]{xcolor}
\usepackage{tikz}
\usetikzlibrary{arrows.meta}
\definecolor{mycolor}{RGB}{128,0,0}
\usetikzlibrary{external}
\begin{document}
\title{Interfacial thermal transport with strong system-bath coupling: A phonon delocalization effect}
\author{Dahai He}
\email[]{dhe@xmu.edu.cn}
\affiliation{Department of Physics and Institute of Theoretical Physics and Astrophysics, Xiamen University, Xiamen 361005, China}
\affiliation{Department of Chemistry, Massachusetts Institute of Technology, 77 Massachusetts Avenue, Cambridge, Massachusetts 02139, USA}
\author{Juzar Thingna}
\affiliation{Complex Systems and Statistical Mechanics, Physics and Materials Science Research Unit, University of Luxembourg, L-1511 Luxembourg, Luxembourg}
\author{Jianshu Cao}
\affiliation{Department of Chemistry, Massachusetts Institute of Technology, 77 Massachusetts Avenue, Cambridge, Massachusetts 02139, USA}
\date{\today}

\begin{abstract}
We study the effect of system-bath coupling strength on quantum thermal transport through the interface of two weakly coupled anharmonic molecular chains using quantum self-consistent phonon approach. The heat current shows a resonant to bi-resonant transition due to the variations in the interfacial coupling and temperature, which is attributed to the delocalization of phonon modes. Delocalization occurs only in the strong system-bath coupling regime and we utilize it to model a thermal rectifier whose ratio can be non-monotonically tuned not only with the intrinsic system parameters but also with the external temperature.
\end{abstract}

\maketitle
\section{Introduction}\label{sec:1}
When characteristic lengths of nanomaterials approach mean free paths of phonons, thermal processes are no longer driven by scattering inside the bulk materials, but driven instead by scattering at heterojunction interfaces. The presence of interfacial thermal resistance hinders energy dissipation inside modern integrated electronic devices, which has become a severe obstacle to their sustainability and integration at nanoscale. Therefore, understanding and manipulating interfacial energy transport in microscopic systems is significant from a fundamental perspective as well as in practical applications~\cite{Li_rev12, Cahill_rev14, Volz_rev16}. In recent years, interfacial thermal transport has been extensively explored in both classical and quantum systems of low-dimensional atomic junctions~\cite{Li05, Hu_SCPT06, He09, Zhang11, Cao15}. However, the effect of strong system anharmonicity in presence of strong system-bath coupling remains an unclarified issue.

In the classical regime, the role of system-bath coupling have been studied in strongly anharmonic systems~\cite{Tolla93,Lepri_rev03}. However, in the quantum regime the role of strong anharmonicity along with strong system-bath coupling have been rarely explored. One of the reasons is perhaps the lack of an effective approach that can access a broad range of parameters. The popular approaches in the quantum regime rely either on the nonequilibrium Greens function (NEGF) method~\cite{Wang_revEPJB08, Wang_revFront14} that deals with evaluating the nonequilibrium correlation functions or the quantum master equation (QME) techniques~\cite{Segal05, Thingna12, Thingna14, Thingna17} that evaluate the reduced density matrix of the system. The NEGF approach can deal with strong system-bath couplings but is unable to capture strong anharmonicity whereas the QME approach deals with the system-bath coupling perturbatively. Other sophisticated approaches based on path-integral techniques or polaron transformation have been restricted mainly to the spin-boson model and have studied the control of heat flux and its geometric properties in the strong system-bath coupling regime~\cite{Segal06, Nicolin11, Wang15, Xu16, Wang17, liu17}.

Recently, a quantum self-consistent phonon theory (QSCPT) has been proposed to study thermal transport through anharmonic quantum systems in a feasible and effective manner~\cite{He16}. The approach incorporates the effects of strong system-bath coupling and strong anharmonicity simultaneously by replacing the anharmonic Hamiltonian with an effective Harmonic one. In this work, we apply the self-consistent phonon approach to a hybrid anharmonic system that couples strongly to two ohmic baths at different temperatures. We are mainly interested in the interfacial transport between two anharmonic segments of our model in the strong system-bath coupling regime. The heat current shows a resonant to bi-resonant transition as a function of the interfacial coupling that is associated to the presence of delocalized phonon modes which persist even due to the presence of strong anharmonicity. Moreover, the presence of anharmonicity aides to observe the transition even as a function of the average bath temperature, which displays the same underlying physics of the delocalization. We further explore thermal rectification in our model and show that it can be non-monotonically controlled via either the system-bath coupling or the average bath temperature.

The paper is organized as follows: In Sec.~\ref{sec:2}, we introduce the anharmonic model and provide a brief overview of the quantum self-consistent phonon approach. In Sec.~\ref{sec:4}, we investigate the heat current and elucidate the resonant to bi-resonant transition along with the underlying mechanism that arises due to the delocalization of phonon modes. Moreover, we construct a special thermal rectifier whose rectification ratio can be fully controlled. Finally, we summarize our main conclusions in Sec.~\ref{sec:5}.
%
\begin{figure}[t!]
\includegraphics[width=\columnwidth]{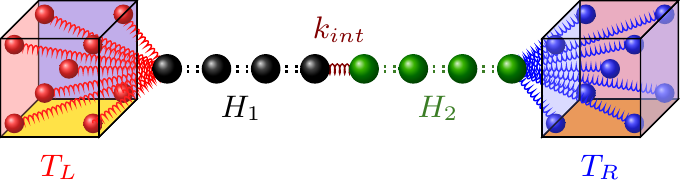}
\caption{\label{fig1}(Color Online) Schematic illustration of the anharmonic model given by Eq.~\eqref{s2-eq-Hamiltonian}. The left and right harmonic baths are at temperatures $T_L$ and $T_R$ respectively. The central system  depicted by Eq.~\eqref{s2-eq-Hs} is a one-dimensional Fermi-Pasta-Ulam-Tsingou $\beta$ chain consisting of two weakly-coupled ($H_1$ and $H_2$) segments.}
\end{figure}
\section{Model and Theory}\label{sec:2}
We consider the Zwanzig-Caldeira-Leggett model~\cite{Zwanzig73, Caldeira83} of dissipation extended to thermal transport, wherein we have two heat baths at different temperatures (as illustrated in Fig.~\ref{fig1}) linearly coupled to a quantum anharmonic chain. The total Hamiltonian for such a closed system reads~\cite{Weiss08},
\begin{eqnarray}\label{s2-eq-Hamiltonian}
H = H_{S} &+& \sum_{l}\frac{P_{l}^2}{2M_{l}}
+\frac{M_{l}\omega_l^2}{2}\left(Q_l-\frac{c_l S_{L}}{M_{l}\omega_l^2}\right)^{2} \nonumber\\
&+&\sum_{r}\frac{P_{r}^2}{2M_{r}}+\frac{M_{r}\omega_r^2}{2}\left(Q_r-\frac{c_r S_{R}}{M_{r}\omega_r^2}\right)^{2},
\end{eqnarray}
where $H_{S}$ describes the system of interest. The bath is a collection of harmonic oscillators with \{$Q_{y}$, $P_{y}$, $M_{y}$, $\omega_{y}$\} describing the position, conjugate momenta, mass, and frequency modes of the two baths ($y=l,r$). The parameter $c_{y}$ is the coupling strength of the $y$th mode of the bath and the system. The system operator $S_{\alpha}$ couples the system to the $\alpha$th bath and in general it can be any system operator or its function. Segregating Eq.~\eqref{s2-eq-Hamiltonian}  into various regions,
\begin{equation}
H=H_{S}+\sum_{\alpha = L,R} (H_{\alpha}+H_{S\alpha}+H_{\alpha}^{RN}),
\end{equation}
where the bath Hamiltonian
\begin{equation}\label{s2-eq-Hamiltonianbath}
H_{\alpha}=\sum_{y}\frac{P_{y}^{2}}{2M_{y}}+\frac{1}{2}M_{y}\omega_{y}^{2}Q_{y}^{2}.
\end{equation}
The interaction Hamiltonian is given by
\begin{equation}\label{s2-eq-Hamiltoniansysbath}
H_{S\alpha}=S_{\alpha}\otimes B_{\alpha},
\end{equation}
where $B_{\alpha}=-\sum_{y}c_{y}Q_{y}$ is the collective bath operator that couples with the system and
\begin{equation}\label{s2-eq-Hamiltonianrn}
H_{\alpha}^{RN}=\frac{S_{\alpha}^2}{2}\sum_{y}\frac{c_{y}^{2}}{M_{y}\omega_{y}^{2}}
\end{equation}
is known as the re-normalization (counter) term~\cite{Weiss08} that is required to ensure homogeneous dissipation for the system particle linked by $S_{\alpha}$. The properties of the bath will be expressed via the spectral density
\begin{equation}
\label{sepctral}
J_{\alpha}(\omega)=\frac{\pi}{2}\sum_{y}\frac{c_{y}^{2}}{M_{y}\omega_{y}}\delta(\omega-\omega_{y}),
\end{equation}
that effectively accounts for the dissipation strength (square of the system-bath coupling strength) and the density of the states of the bath. In the above equations $\alpha = L;~y=l$ corresponds to the left bath and $\alpha = R;~y=r$ corresponds to the right bath.

In this work the system comprises of two weakly coupled one-dimensional anharmonic segments whose Hamiltonian is described by
\begin{equation} \label{s2-eq-Hs}
H_{S} = H_{1}+\frac{k_{int}}{2} (x_{N/2+1}-x_{N/2})^2+H_{2}.
\end{equation}
Each segment will be the archetypal Fermi-Past-Ulam-Tsingou (FPUT) $\beta$ model whose Hamiltonian
\begin{equation}
H_{j} = \sum^{N_{j}^{h}}_{n=N_{j}^{l}} \frac {p_n^2}{2m} +V_{j}(x_{n+1}-x_{n}),
\end{equation}
where $j=1$ ($N_1^{l}=1$; $N_1^{h}=N/2$) is the left segment connected only via particle $1$ to the left bath, i.e., $S_{L} = x_1$ and $j=2$ ($N_2^{l}=N/2+1$; $N_2^{h}=N$) corresponds to the right segment with $S_{R}=x_N$ [refer Eq.~(\ref{s2-eq-Hamiltoniansysbath})]. The anharmonic potential in each segment is given by
\begin{equation} \label{s1-eq-int-potential}
V_{j}(x)=\frac{1}{2}k_{j}x^2+\frac{1}{4}\lambda_{j}x^4.
\end{equation}
Since the coupling between these two segments $k_{int}$ is considered weak it acts like an impurity that will lead to phonon scattering. Throughout this work we will use the term interface for the coupling between the two anharmonic chains. Importantly, the weak interfacial coupling between the two anharmonic segments will ensure that each segment attains approximately the same temperature as that of the bath it is connected too. This allows us to calculate the heat current through the anharmonic quantum system using quantum self-consistent phonon theory (QSCPT)~\cite{Feynman86, He08a, He16}.

The key idea behind the QSCPT is to assume a trial harmonic Hamiltonian $H_{S}^{eff}$ with \emph{undetermined} parameters using which we can obtain canonical averages analytically. In a non-equilibrium set-up, the use of canonical averages is strictly valid if there is no temperature gradient across the system. This condition holds if the system is a molecular junction comprising of only a few atoms or when the temperature of the baths is in the linear response regime. For long chains beyond linear response, the condition trivially holds for an ordered harmonic system since it has a flat temperature profile \cite{Dhar_rev08}. The non-trivial scenario where this assumption holds is the two-segment model studied herein where due to the weak interfacial coupling each segment can be assumed to attain a local equilibrium with the temperature of its respective bath. Furthermore, we ensure that each segment comprises only of a few atoms such that the mean-free path of the phonons, which is typically hundreds of atoms \cite{Liu2014, Saaskilahti2015}, is much longer than the segment length. Hence, QSCPT could be applied independently to the two segments without the need to deal with non-equilibrium averages \cite{Zhang13}.

The trial harmonic Hamiltonian for the system is chosen to be,
\begin{equation}
\label{s3-eq-Hamiltonianeff}
H_{S}^{eff} = H_{1}^{eff}+\frac{k_{int}}{2} (x_{N/2+1}-x_{N/2})^2+H_{2}^{eff},
\end{equation}
where the effective Hamiltonian of the two segments take the form
\begin{equation}\label{s3-eq-H1eff}
H_{j}^{eff} = \sum^{N_{j}^h}_{n=N_{j}^{l}} \frac {p_n^2}{2m} +\frac {f_{j}}{2} (x_{n+1}-x_{n})^2.
\end{equation}
The trial parameters are obtained in a self-consistent way by minimizing the upper bound of free energy, given by the Feynman-Jensen inequality~\cite{Feynman98}. Thus, the set of self-consistent equations obtained for each segment \cite{He16} read
 \begin{eqnarray}
\label{eq:fc}
f_{j}&=& k_{j}+\frac{3\lambda_{j}}{Nm}\sum_{p}
\frac{4\sin^{2}\left(\frac{2p\pi}{N}\right)}{\omega_{j}(p)}\coth\left[\frac{\beta_{j}\omega_{j}(p)}{2}\right], \nonumber \\
\label{eq:wp}
\omega_{j}^{2}(p)&=&\frac{4f_{j}}{m}\sin^{2}\left(\frac{2p\pi}{N}\right),
\end{eqnarray}
where $\beta_{j}=[k_{B}T_{L (R)}]^{-1}$ for $j=1 (2)$.

Using the effective Hamiltonian Eq.~\eqref{s3-eq-Hamiltonianeff} we can obtain the steady-state heat current \cite{note1} through the system given by the Landauer-like formula~\cite{He16},
\begin{equation}\label{s3-eq-current}
I_{L}=-I_{R}=\frac{1}{2\pi}\int_{0}^{\infty} d\omega\omega \widetilde{\mathcal{T}}(\omega)[n_{L}-n_{R}].
\end{equation}
The transmission function $\widetilde{\mathcal{T}}(\omega)$ is given by the Caroli formula \cite{Caroli71},
\begin{equation}
\widetilde{\mathcal{T}}(\omega)=\mathrm{Tr}(\mathbf{G}^{r}\mathbf{\Gamma}_{L}\mathbf{G}^{a}\mathbf{\Gamma}_{R}),
\label{s3-eq-caroli}
\end{equation}
where
\begin{eqnarray}
\label{s3-eq-Greensfunc}
\mathbf{G}^r&=&\left[ m\omega^2\mathbf{I}-\widetilde{\mathbf{K}}-\mathbf{\Sigma}^{r}\right]^{-1};\quad \mathbf{G}^a=(\mathbf{G}^r)^{\dag},\\
\label{s3-eq-Gamma}
\mathbf{\Gamma}_{\alpha}&=&-2\mathrm{Im}(\mathbf{\Sigma}_{\alpha}^{r});\quad n_{\alpha}=\left(\exp{(\beta_{\alpha}\omega)}-1\right)^{-1}.
\end{eqnarray}
Here $\mathbf{I}$ is the identity matrix and $\mathbf{\Sigma}^{r}=\mathbf{\Sigma}_{L}^{r}+\mathbf{\Sigma}_{R}^{r}$ is sum of the retarded self-energy of the left $\mathbf{\Sigma}_{L}^{r}$ and right $\mathbf{\Sigma}_{R}^{r}$ baths with two nonzero elements $\mathbf{\Sigma}^{r}_{1,1}=\bar{\Sigma}_{L}^{r}$ and $\mathbf{\Sigma}^{r}_{N,N}=\bar{\Sigma}_{R}^{r}$ given by,
\begin{eqnarray}
\bar{\Sigma}_{\alpha}^{r}(\omega)&=&\frac{1}{\pi}\mathrm{P}\int_{-\infty}^{\infty}\frac{J_{\alpha}(\omega')d\omega'}{\omega-\omega'}-iJ_{\alpha}(\omega) \nonumber \\
&&+\frac{2}{\pi}\int_{0}^{\infty} d\omega'\frac{J_{\alpha}(\omega')}{\omega'},
\end{eqnarray}
where the last frequency independent term in the self-energy arises due to the re-normalization of the Hamiltonian [Eq.~(\ref{s2-eq-Hamiltonianrn})]. Above, $n_{\alpha}$ is the phonon distribution of the $\alpha$th heat bath and $\widetilde{\mathbf{K}}$ is the effective tridiagonal force matrix for the harmonic Hamiltonian Eq.~\eqref{s3-eq-Hamiltonianeff}. Since the trial parameters $f_{1}$ and $f_{2}$ are temperature dependent, the transmission function $\widetilde{\mathcal{T}}(\omega)$ depends on temperature of both baths for anharmonic nonequilibrium systems.

The model is completed by specifying the spectral density of heat baths see Eq.~(\ref{sepctral}). Without loss of generality, we assume that both the left and right baths have the same spectral density, i.e., $J_{L}(\omega) = J_{R}(\omega) = J(\omega)$ and use the Ohmic spectral density with a Lorentz-Drude cutoff $\omega_c$ given by,
\begin{equation}
J(\omega)=\frac{\gamma m \omega}{1+(\omega/\omega_{c})^{2}}.
\end{equation}
The parameter $\gamma$ is the Stokesian damping coefficient which dictates the strength of dissipation. The dissipation strength is proportional to the sum of squares of individual coupling strengths between the bath oscillators and the system, i.e., $\gamma \propto \sum_n c_n^2$ and hence it serves as a measure of the system-bath coupling strength. Thus, for the Ohmic spectral density with Lorentz-Drude cutoff the non-zero elements of the retarded self-energy of the $\alpha$th is given by
\begin{equation}
\bar{\Sigma}_{\alpha}^{r}(\omega)=J(\omega)\left[\frac{\omega}{\omega_{c}}-i\right].
\end{equation}

In the next section we now explore the effects of strong system-bath coupling on nonequilibrium anharmonic systems and elucidate how in such systems we can take advantage of a strong coupling to obtain practical applications such as rectification. Throughout this work we have set the Planck constant $\hbar=1$ and the Boltzmann constant $k_B=1$.
\section{Results and discussions}\label{sec:4}
%
\begin{figure}[t!]
\includegraphics[width=\columnwidth]{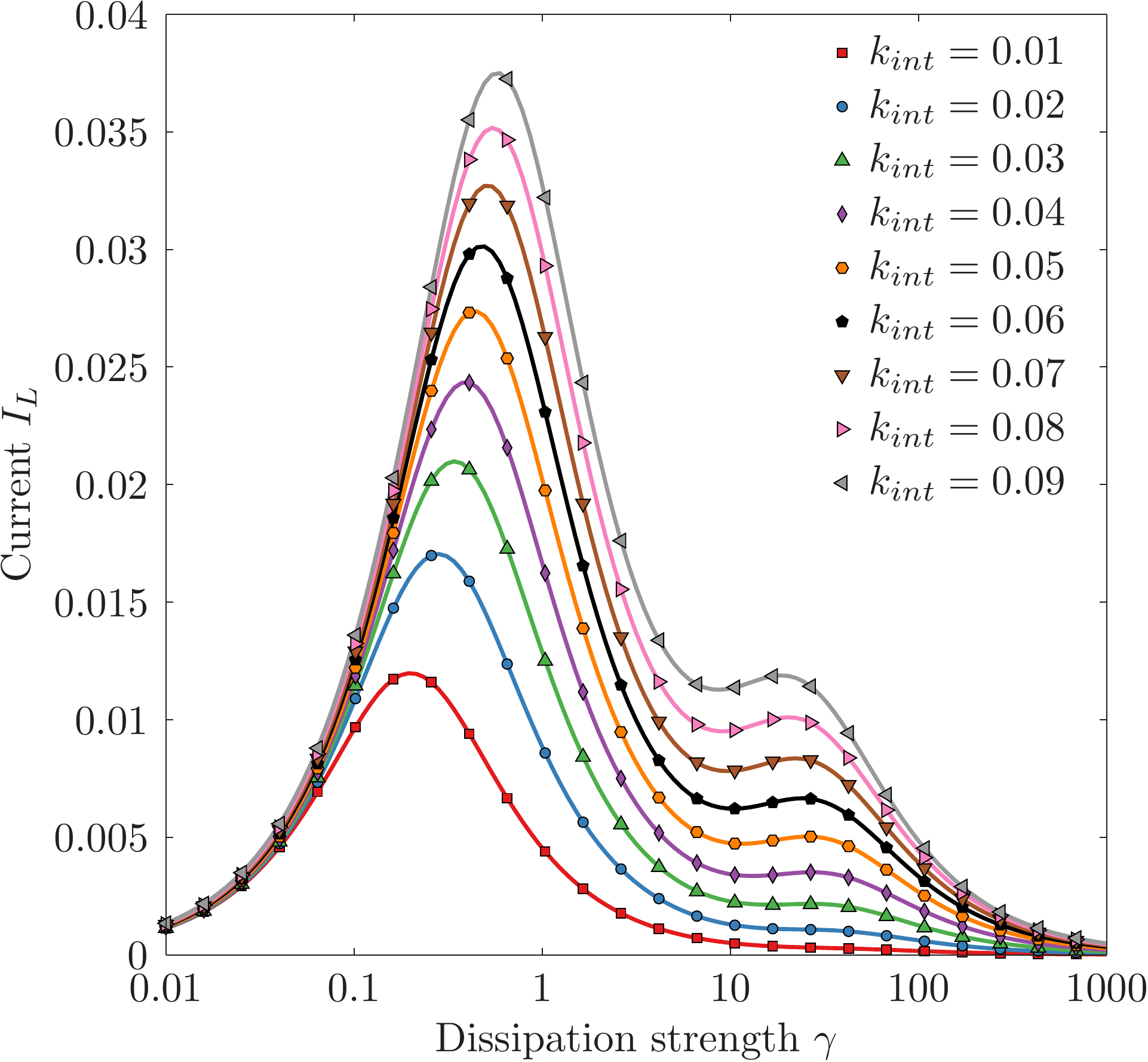}
\caption{\label{fig2}(Color Online) Current $I_{L}$ as a function of dissipation strength $\gamma$ for various strengths of $k_{int}$. Common parameters are: $m=$ 1, $k_{1}=k_{2}=$ 2, $\lambda_{1}=\lambda_{2}=$ 1, $\omega_{c}=$ 10, $T=0.8$, and $\Delta=0.6$. The left bath temperature $T_{L} = T\left(1+\Delta\right)$, whereas the right bath temperature $T_{R} = T\left(1-\Delta\right)$.}
\end{figure}
\subsection{Delocalized phonon modes and interfacial thermal transport}
Using the QSCPT described in the previous section we evaluate the heat current [Eq.~\eqref{s3-eq-current}] for a system of total length $N=8$ [Eq.~\eqref{s2-eq-Hs}], kept fixed throughout, and focus on the strong system-bath coupling effects. We find that the maximum resonant current is obtained in the moderate system-bath coupling regime in the vicinity of $\gamma = 1$ and decreases beyond the peak because the dissipative effects dominate \cite{Lepri_rev03, Wang15, Wang17} as shown in Fig.~\ref{fig2}. For values of interfacial coupling $k_{int} \geq 0.03$ we see a second off-resonant peak emerging in the current for larger values of the dissipation strength $\gamma$.

\begin{figure}[t!]
\includegraphics[width=\columnwidth]{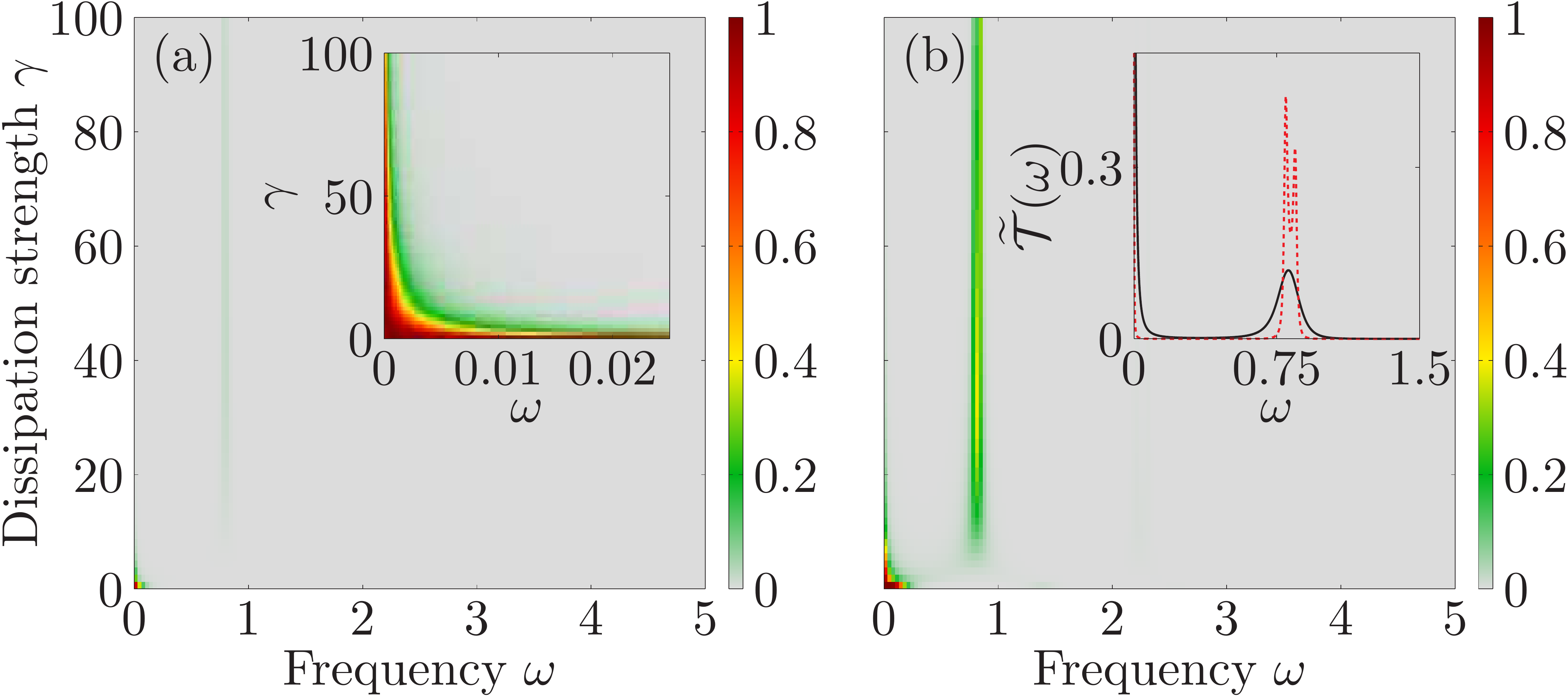}
\caption{\label{fig3}(Color Online) The color map of the transmission function $\widetilde{\mathcal{T}}(\omega)$ for $k_{int}=0.01$ [panel (a)] and $k_{int}=0.05$ [panel (b)]. All others parameters are the same as that for Fig.~\ref{fig2}. The inset in panel (b) is the transmission for $\gamma=10$ (black solid line) and $\gamma=80$ (red dashed line). The transmission for $\omega = 0$ for all values of $k_{int}$ and $\gamma$ is nearly 1 and decays quickly to zero as seen in the inset in panel (a).}
\end{figure}

In order to understand the emergence of the off-resonant peak we study the transmission function $\widetilde{T}(\omega)$ as shown in Fig.~\ref{fig3}. The transmission for $k_{int} < 0.03$ shows a peak only at $\omega = 0$ and is nearly zero everywhere else [Fig.~\ref{fig3}(a)] whereas for $k_{int} \geq 0.03$ the transmission shows the first peak at $\omega = 0$ and another minor peak close to $\omega = 1$ [Fig.~\ref{fig3}(b)]. Since we don't have any on-site potential that breaks momentum conservation (translational invariance), the long wavelength mode $\omega \rightarrow 0$ has the maximum transmission $=1$ [see Fig.~\ref{fig3}(a) inset] since it views the finite system as a single massive atom without any intricate details, e.g. the interface \cite{Zhang11}.

The intriguing second peak in the transmission for $k_{int}\geq 0.03$ is a result of Fabry-P\'{e}rot-like interference. Such interference patterns have been observed in ordered harmonic systems without an interface where the boundaries to the baths act like partially reflecting surfaces for the phonon modes~\cite{Hyldgaard04, Hopkins2009, Hu2010, Whitney2015}. For such systems the interference pattern leads to delocalized phonon modes (seen as peaks in the transmission) that resonate with the system-phonon modes (normal modes of the effective force matrix $\widetilde{\mathbf{K}}$). In this work, we observe the survival of such an interference pattern in strongly anharmonic systems as seen in Fig.~\ref{fig3}. The strong delocalization observed in ordered harmonic chains becomes weak due to the presence of an additional weak interface $k_{int}$ that acts as an impurity whose presence would be reflected in all non-zero frequency modes.

The strength of $k_{int}$ dictates the weight with which the modes get delocalized and hence for very weak $k_{int} < 0.03$ we find that the transmission is nearly zero for all non-zero frequency modes [see Fig.~\ref{fig3}(a) with inset focusing on the zero frequency for $k_{int} = 0.01$]. Beyond this value ($k_{int} \geq 0.03$) in the strong system-bath coupling regime $\gamma \geq 10$, we find the lowest non-zero frequency mode gets delocalized and starts transmitting. A weak $k_{int}$ allows only the lowest frequency phonon mode of the system to transmit, hence we see its strong presence in the transmission [Fig.~\ref{fig3}(b)]. Furthermore, in the strong system-bath coupling regime since the delocalization frequency of this mode depends only on the system-phonon modes and interfacial coupling, we don't find a strong dependence on the dissipation strength $\gamma$ as seen in Fig.~\ref{fig3}(b) inset.

Thus, overall increasing the interfacial coupling opens up more transmission channels that would eventually become equal to the system-phonon modes when both the interfacial coupling and the system-bath coupling are strong. The opening of these extra channels leads to an increase in the heat current and moreover leads to the double peaked structure as seen in Fig.~\ref{fig2}. We note here that if more phonon modes get delocalized as the $k_{int}$ is further increased we don't see an increase in the number of peaks in the current but rather that the peaks become more pronounced. This implies that in the strong system-bath coupling regime there are two distinct physical mechanisms that correspond to the bi-resonant behavior in the heat current. -- One stemming from the long-wavelength contribution (zero frequency) that ignores the intricate details of the system, like the presence of a weak interface. -- Second due to short-wavelength modes (non-zero frequency) that take into account the system details.
\begin{figure}[t!]
\includegraphics[width=\columnwidth]{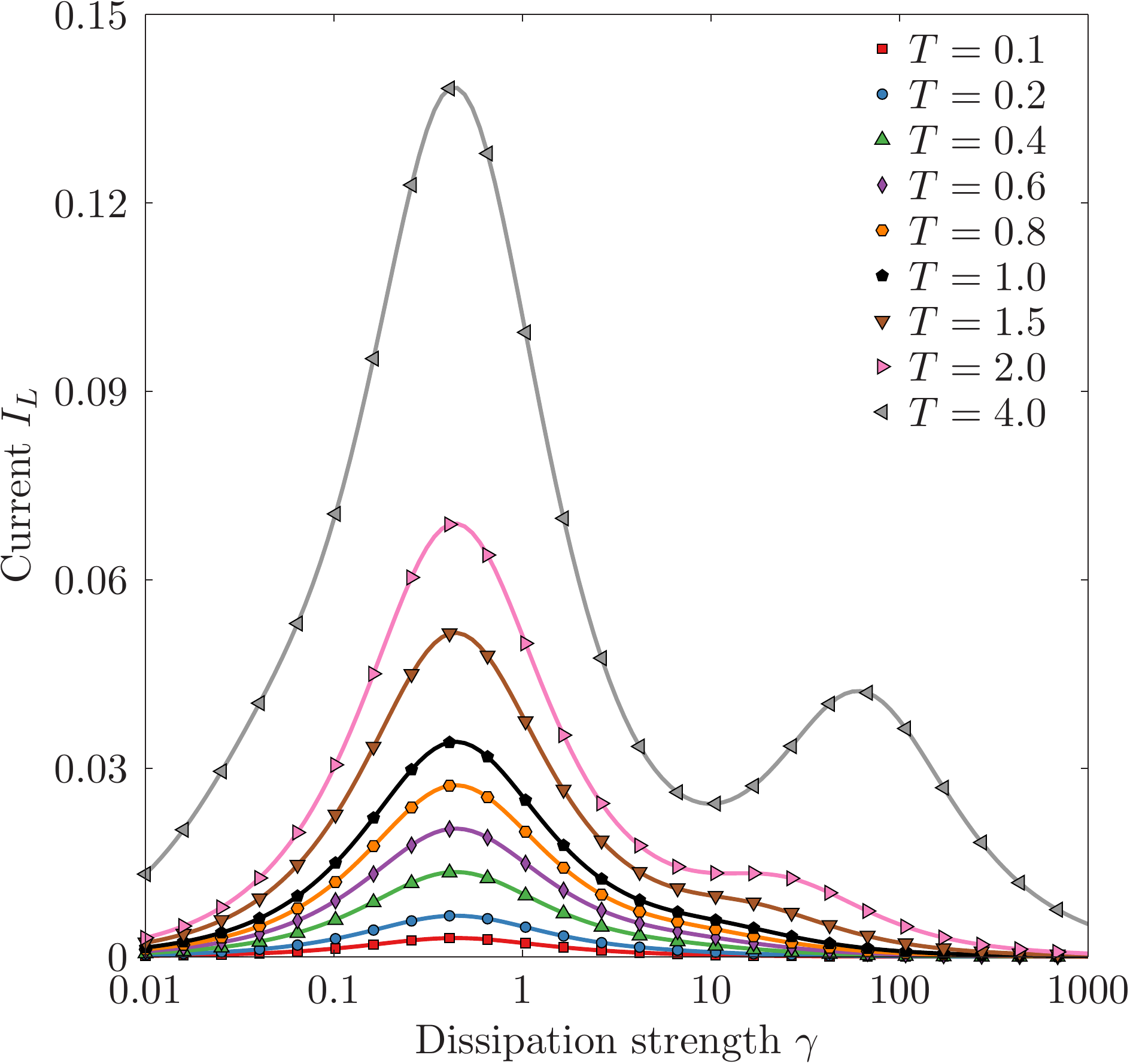}
\caption{\label{fig4}(Color Online) Current $I_{L}$ as a function of the dissipation strength $\gamma$ for various average temperatures. Parameters used for the calculations are: $m=$ 1, $k_{1}=1, k_{2}=2$, $\lambda_{1}=0.5, \lambda_{2}=$ 1, $k_{int}=$ 0.05, and $\omega_{c}=$ 10. The left bath temperature $T_{L} = T\left(1+\Delta\right)$, whereas the right bath temperature $T_{R} = T\left(1-\Delta\right)$ with $\Delta$ fixed as 0.6.}
\end{figure}

Next, we study the effect of temperature $T=(T_L+T_R)/2$ in our model. According to QSCPT, the presence of anharmonicity causes the transmission function to be temperature dependent [refer Eqs.~(\ref{s3-eq-caroli}) and~(\ref{eq:fc})]. Thus, a variation in temperature can be understood as an effective change in parameters of the two anharmonic segments according to Eqs.~(\ref{s3-eq-Hamiltonianeff}), (\ref{s3-eq-H1eff}), and (\ref{eq:fc}). Therefore, one intuitively expects that the heat current, when temperature is varied, should have similar trends as compared to the case when the interfacial coupling is changed. This is because in the former case we keep the interfacial coupling constant and effectively change the system parameters whereas in the latter the effective system parameters are held fixed.

\begin{figure}[t!]
\includegraphics[width=\columnwidth]{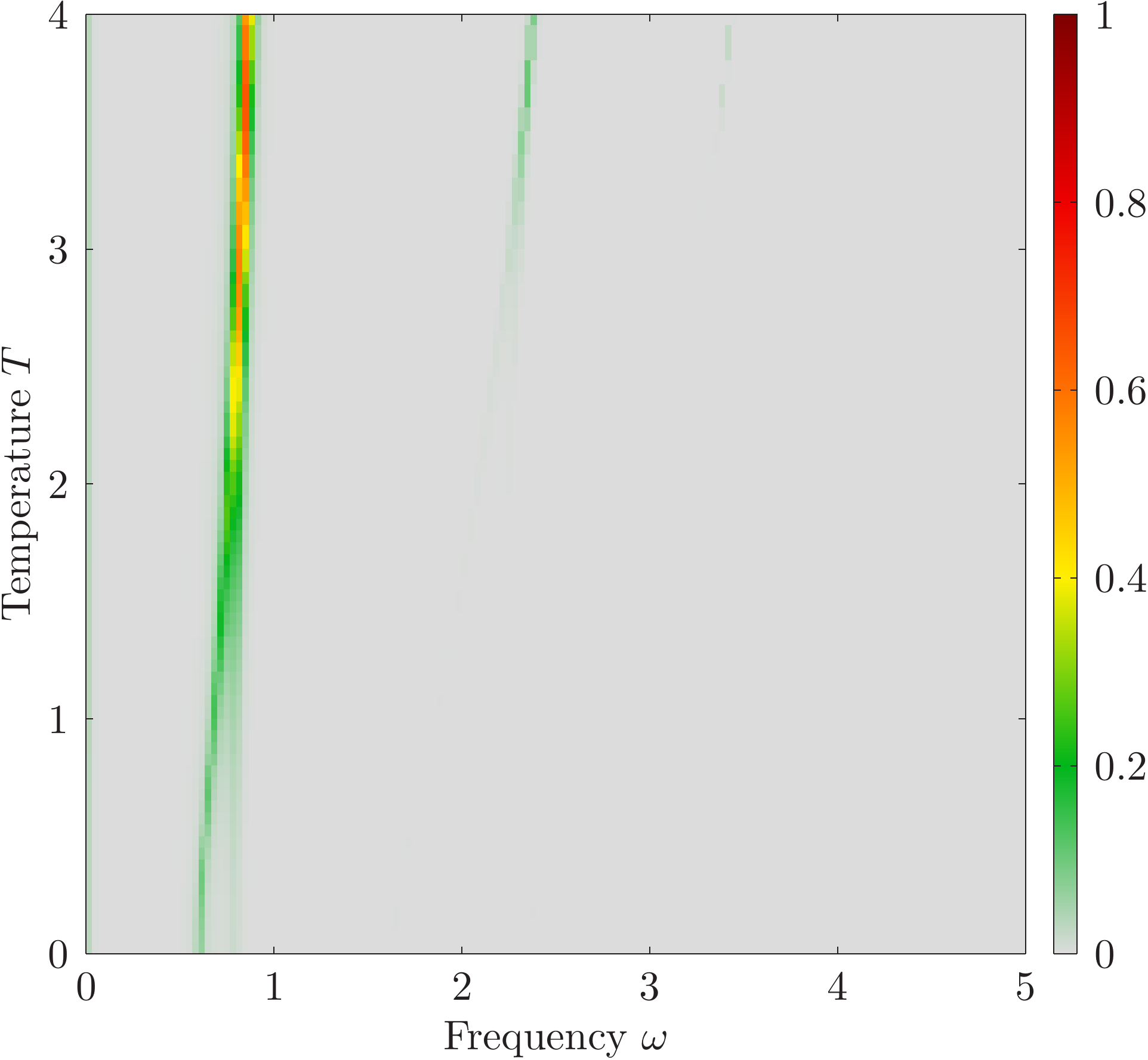}
\caption{\label{fig5}(Color Online) The color map of transmission function $\widetilde{\mathcal{T}}(\omega)$ for  $\gamma=30$. All others parameters are the same as that for Fig.~\ref{fig4}.}
\end{figure}

Figure~\ref{fig4} displays the heat current for various values of temperature $T$ and similar to the interfacial coupling variation we find a resonant to bi-resonant transition. The low temperature regime displays a resonant behavior whereas beyond the critical temperature $T_c=1.5$ we observe a bi-resonant structure in the heat current. The resonant peak remains independent of temperature since the underlying phonon modes contributing to this peak are the zero frequency modes. The off-resonant peak is due to the delocalization of phonon modes as seen from the transmission function in Fig.~\ref{fig5} and interestingly the temperature variation also leads to shift in delocalization frequency. This shift in the delocalization frequency is reflected as a shift in the off-resonant peak of the heat current as a function of dissipation strength $\gamma$ as shown in Fig.~\ref{fig4}. Thus, the resonant to bi-resonant transition in the heat current can be observed not only by changing the intrinsic system properties, like the interfacial coupling, but also by varying the temperature of the leads. This transition requires the delocalization of non-zero frequency phonon modes and the temperature variation can only be observed in highly anharmonic systems.
\subsection{Thermal rectification}
Analogous to the electronic diodes, heat diodes rely on thermal rectification that has been well explored in anharmonic systems and nanostructures~\cite{Terraneo02,Li04,Hu_SCPT06,Zeng08,Li_rev12}. Thermal rectification is the asymmetric flow of heat when the temperatures of the two baths are interchanged, hence its presence leads to a directional heat flow. In case of our asymmetric two-segment system with a weak interfacial coupling, heat is exchanged efficiently when the system-phonon frequency in one segment (normal mode of the partial one-segment force matrix $\widetilde{\mathbf{K}}$) matches that in the other segment. In other words, only phonons with frequency inside the overlapping phonon bands, i.e. frequency range of allowed phonons~\cite{Terraneo02}, of the two segments can be transported through the interface.

\begin{figure}[t!]
\includegraphics[width=\columnwidth]{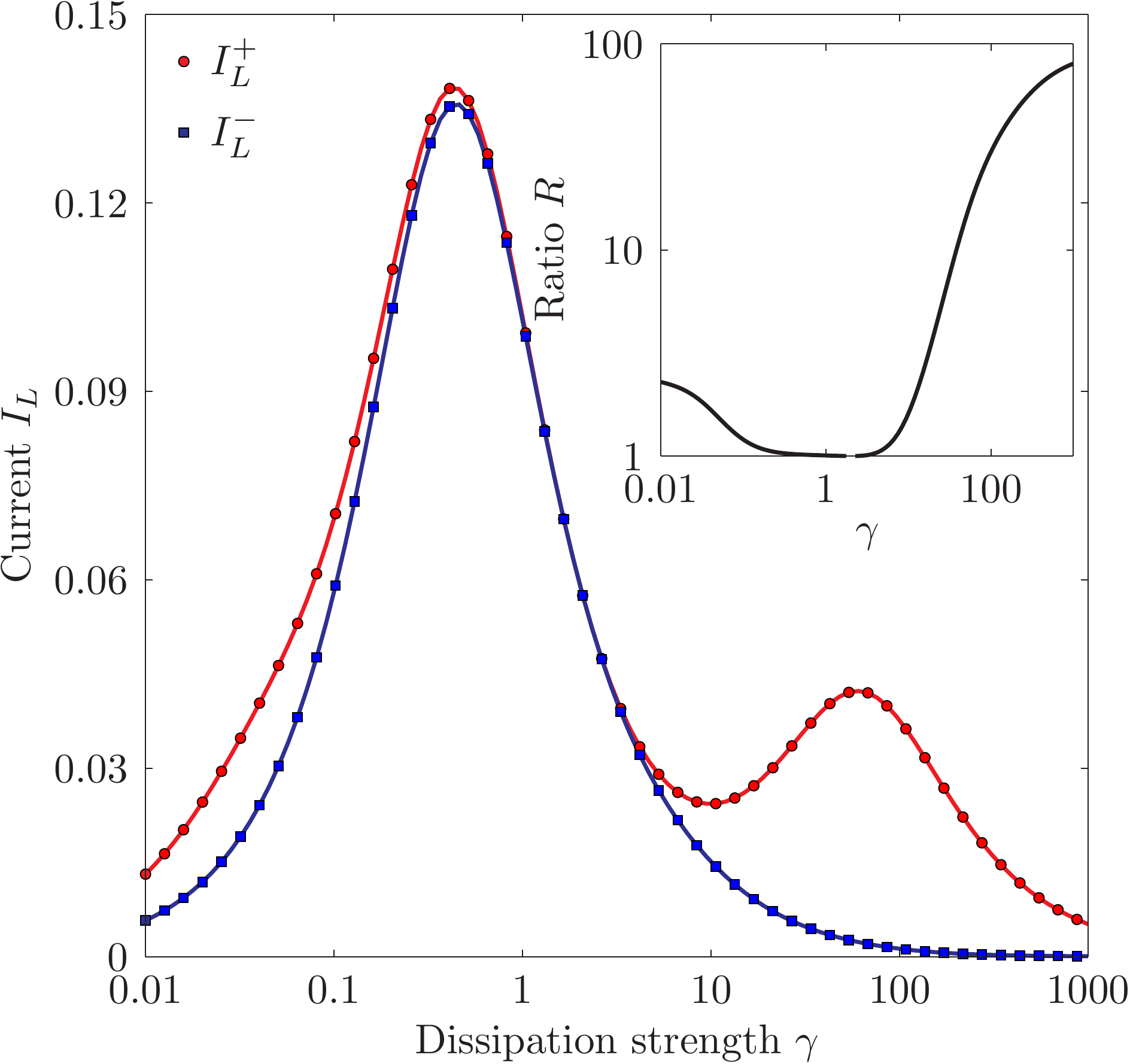}
\caption{\label{fig6}(Color Online) Forward Current $I_{L}^{+}$ and backward current $I_{L}^{-}$ as a function of the dissipation strength $\gamma$. Inset gives the rectification ratio $R = I_{L}^{+}/I_{L}^{-}$ as a function of $\gamma$, in which one can see a plateau $R\approx1$ (`off' state) around $\gamma = 1$. Parameters used for the calculation are: $m=$1, $k_{1}=$1, $k_{2}=2$, $\lambda_{1}=0.5$, $\lambda_{2}=1$, $k_{int}=$ 0.05, $\omega_{c}=$10, $T=$4, and $\Delta=0.6$.}
\end{figure}

According to QSCPT, the presence of anharmonicity results in temperature dependence of the system-phonon frequencies [see Eq.~(\ref{eq:fc})]. When the baths are interchanged the local equilibrium temperature of each segment differs causing the effective force matrix of each segment to change. Thus, the effect of interchanging the baths causes the effective system parameters to change and hence the underlying effect is similar to that of varying temperature as explained in the previous section. We demonstrate the consequence of this phenomenon in Fig.~\ref{fig6} wherein the forward current $I_L^+$ (when the left segment is connected to the hot bath) has only a single peak whereas the backward current $I_L^-$ (when the right segment is connected to the hot bath) has two peaks. The presence of the bi-resonant structure for the backward current is due to the delocalization of phonon modes and in this case turns out to be advantageous to have a extremely high rectification ratio $R\equiv I_{L}^{+}/I_{L}^{-}$ as seen in Fig.~\ref{fig6} inset.

\begin{figure}[t!]
\includegraphics[width=\columnwidth]{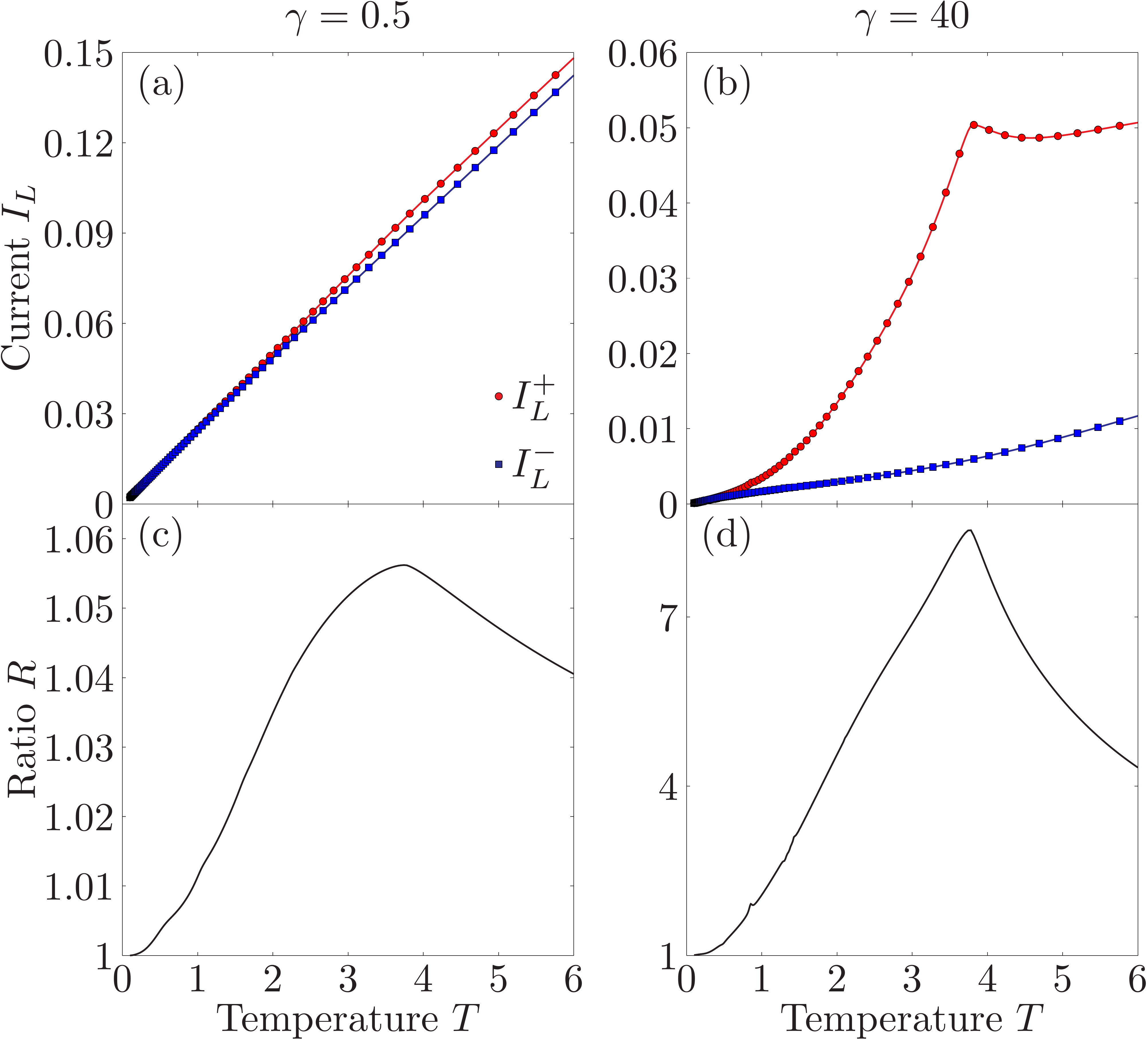}
\caption{\label{fig7}(Color Online) Forward Current $I_{L}^{+}$ and backward current $I_{L}^{-}$ as a function of the temperature $T=(T_L+T_R)/2$ in the weak [panel (a)] and strong [panel (b)] system-bath coupling regimes. Panels (c) and (d) depict the corresponding rectification ratio $R = I_{L}^{+}/I_{L}^{-}$ as a function of $T$ that shows a drastic enhancement in the strong system-bath coupling regime [panel (d)]. Other parameters used are same as Fig.~\ref{fig6}.}
\end{figure}

Moreover, the underlying physics of delocalization can also help tune the rectification ratio as a function of the average temperature $T$ as seen in Fig.~\ref{fig7}. In the weak system-bath coupling regime $\gamma=0.5$ the forward and backward currents are nearly the same [Fig.~\ref{fig7}(a)] leading to a relatively minor enhancement in the rectification ratio [Fig.~\ref{fig7}(c)]. Whereas, in the strong system-bath coupling regime we see a significant enhancement [Fig.~\ref{fig7}(b) and (d)], due to the presence of delocalized modes. In a nutshell, the presence or absence of delocalized modes occurs only in the strong system-bath coupling regime and depends significantly on the effective force matrix of the two segments and the interfacial coupling. Since the phenomenon of rectification relies on the interchange of the two heat baths it affects the effective force matrix via the phonon frequencies of the two segments, due to their local equilibrium temperature, as easily seen from the QSCPT. This overall affects the forward and backward currents leading to a significant enhancement of the rectification ratio, not only as a function of the dissipation strength $\gamma$ but also with respect to the easily tuned external parameter, i.e, the average temperature $T$.
%
\section{Summary}\label{sec:5}
In this paper we investigated the effects of strong system-bath coupling, characterized by dissipation strength $\gamma$, on interfacial quantum thermal transport through two anharmonic segments. We employ the quantum self-consistent phonon theory (QSCPT) to turn the anharmonic Hamiltonian into an effective Harmonic one whose parameters depend on the temperature of the baths. In the extremely weak interfacial coupling regime, we find that only the long wavelength mode is able to transmit and the overall thermal transport properties are independent of the local system details. Increasing the interfacial coupling opens up channels of transmission that appear as delocalized phonon modes in the strong system-bath coupling regime. These extra channels result in an off-resonant peak in the heat current that could be enhanced or suppressed not only as a function of the interfacial coupling but also the temperature. The temperature dependence is a direct consequence of QSCPT because a change in temperature implies a change in effective system parameters of each segment that is equivalent to the weakening or strengthening of the interfacial coupling.

Furthermore, using the off-resonant peak we demonstrated that such a system could be an effective thermal rectifier. Such a rectifier could be turned `on' ($R \gg 1$) or `off' ($R=1$) by either varying the dissipation strength or the temperature. Even though the temperature is an experimentally controllable parameter, the advances in nanotechnology have made the tuning of dissipation strength tunable either by using organic materials to increase the interfacial adhesions~\cite{Kaur14} or by applying high pressure to stiffen the interfacial bonding~\cite{Hohensee15}. Our results provide a stepping stone towards fully understanding the effects of strong system-bath coupling in highly anharmonic systems and their possible utilization in manipulating interfacial thermal transport in low-dimensional nanodevices.

\begin{acknowledgments}
We acknowledge the helpful discussions with Jian-Sheng Wang and J. Mendoza. D.H. is supported by NSFC of China (Grant Nos. 11675133 and 11335006), NSF of Fujian Province (No. 2016J01036), and China Scholarship Council. J.C acknowledges support from the NSF (Grant No. CHE1112825).
\end{acknowledgments}

\begin{thebibliography}{44}
\bibitem [{not()}]{note1}%
  \BibitemOpen
  \href@noop {} {\ }\bibinfo {note} {Since we don't have an external
  time-dependent driving or a flow of particles, the heat current is equivalent
  to the energy current, i.e., $I_{\alpha}=-\langle dH_{\alpha}/dt\rangle$,
  with the brackets denoting a nonequilibrium average.}\BibitemShut {Stop}%
\end{thebibliography}
%
\end{document}